\documentclass[english,a4paper]{article}
\usepackage[T1]{fontenc}
\usepackage[cp1250]{inputenc}
\usepackage{amssymb}
\usepackage{hyperref}
\usepackage{amsfonts}
\usepackage{graphicx}
\usepackage{revsymb}
\usepackage{dcolumn}
\usepackage{bm}

\def\openone{\leavevmode\hbox{\small1\kern-3.3pt\normalsize1}}
\usepackage[usenames,dvipsnames]{color}

\begin{document}
%
%
%
\title{Control of molecular dynamics with zero-area fields: Application to molecular orientation and photofragmentation}
\author{D. Sugny\footnote{Laboratoire Interdisciplinaire
Carnot de Bourgogne (ICB), UMR 5209 CNRS-Universit\'e de
Bourgogne, 9 Av. A. Savary, BP 47 870, F-21078 DIJON Cedex,
FRANCE, dominique.sugny@u-bourgogne.fr}, St\'ephane Vranckx\footnote{Service de Chimie Quantique et Photophysique, CP 160/09 Universit\'e Libre de Bruxelles, B-1050 Brussels, Belgium; Laboratoire de Chimie Physique (LCP), UMR 8000 CNRS-Universit\'e Paris-Sud, Orsay, France}, Mamadou Ndong\footnote{Laboratoire Interdisciplinaire Carnot de Bourgogne,
  ICB UMR 6303 CNRS-Universit\'e de Bourgogne, 9 Avenue A. Savary, BP 47 870 F-21078 Dijon cedex, France}, \\
  Nathalie Vaeck\footnote{Service de Chimie Quantique et Photophysique, CP 160/09 Universit\'e Libre de Bruxelles, B-1050 Brussels, Belgium}, Osman Atabek\footnote{Institut des Sciences Mol\'eculaires d'Orsay (ISMO), B\^at. 350 UMR 8214 CNRS-Universit\'e Paris-Sud, Orsay, France}, Mich\`ele Desouter-Lecomte\footnote{Laboratoire de Chimie Physique (LCP), UMR 8000 CNRS-Universit\'e Paris-Sud, Orsay, France; D\'epartement de Chimie, Universit\'e de Li\`ege, B\^at. B6c, Sart Tilman, B-4000 Li\`ege, Belgium}}

\maketitle

\begin{abstract}
The constraint of time-integrated zero-area on the laser field is a fundamental, both theoretical and experimental requirement in the control of molecular dynamics. By using techniques of local and optimal control theory, we show how to enforce this constraint on two benchmark control problems, namely molecular orientation and photofragmentation. The origin and the physical implications on the dynamics of this zero-area control field are discussed.
\end{abstract}



\section{Introduction}
\label{sec:intro}

Quantum control is aimed at designing external pulses in order to achieve efficient transfers between the states of the quantum system under study \cite{rice,tannorbook,shapiro}. This task is crucial in atomic and molecular physics, and has many applications extending from photochemistry to quantum computation. Quantum control has attracted attention among the physics and chemistry communities \cite{kosloff}, but also in applied mathematics for the development of new theoretical methods. In this context, optimal control theory (OCT) can be viewed as the most accomplished way of designing control fields \cite{pont,revuealessandro,khaneja,lapertglaser}. Several modifications of the standard optimal control algorithms have been brought forward to account for experimental constraints \cite{brif,gross,maday,reich}, such as the non linear interaction of the system with the field \cite{ohtsukinonlinear,lapertalgo}, the question of spectral constraints \cite{viviespectrum,lapertspectrum,kochspectrum} and the robustness with respect to one or several model parameters \cite{grapeino,damping,rabitzino,ferrini}. Recently, we have shown how optimal control strategies can be extended to enforce the constraint of time-integrated zero-area on the control field \cite{JMO}. This constraint is a fundamental requirement in laser physics, as shown in different experimental and theoretical studies \cite{ortigoso,liao,you,timeorientation,JMO,shunew}. Basically this effect can be related to the fact that the dc component of the control field is not a solution of Maxwell's equation. We refer the reader to the first part of the paper for a complete discussion. Note that this point is particularly crucial in the THz regime, or with laser pulses accommodating only few optical cycles \cite{THZ1,THZ2,THZ3,THZ4,THZ5}. The corresponding laser sources are by now commonly used in quantum control \cite{THZcontrol1,THZcontrol2}. Up to now, the majority of the theoretical papers on quantum control does not consider this zero-area requirement \cite{brif,kurosaki}, which may lead to non-physical electromagnetic control fields and is problematic in view of experimental implementations. In addition, imposing such a constraint to the control scheme may force the optimization algorithm to reach more efficient external fields achieving better transfer. These arguments show the importance of the methods and of the results presented in this work, in particular to fill the existing gap between theory and experiment.

Our preliminary study on the subject \cite{JMO} is a methodology-oriented paper, focusing mainly on the technical aspects of the optimization algorithms (local and optimal approaches) as briefly illustrated on two typical molecular processes (orientation and dissociation). The present paper thoroughly expands this initial work \cite{JMO} by providing an extended numerical investigation and a detailed physical analysis of the dynamics of two specific molecular systems. The article is organized as follows. The physical origin of the time-integrated zero-area constraint is presented in Sec.\ref{sec2}. For completeness the principles of the optimization algorithms are also briefly outlined, full technical details being referred to Ref.\cite{JMO}. Section \ref{sec3} focuses on the control of molecular orientation, with the CO molecule as an illustrative example. Section \ref{sec4} is devoted to the dissociation of HeH$^+$ involving the control of a given fragmentation channel. Conclusions and prospective views are given in a final section \ref{sec5}.
\section{Molecular dynamics controlled by zero-area electromagnetic fields}\label{sec2}
\subsection{Origin of the zero-area constraint in molecular physics}
The zero-area constraint for laser pulses although well-known, is rarely given a thorough and clear argumentation. For completeness and pedagogical purposes, this section is devoted to the presentation of such a proof. The body of the proof is made of two parts: The calculation of the time integrated electromagnetic field and the physical interpretation of the result.

We refer to the spatio-temporal electric field amplitude $E(t,\vec r)$ and its Fourier transform $\hat{E}(\omega,\vec k)$, where the corresponding Fourier conjugate variables are time $t$ and frequency $\omega$, on the one hand, and space $\vec r$ and momentum $\vec k$ vectors, on the other hand. These quantities are related through:
\begin{equation}
E(t, \vec r)=\int d\vec k \int d\omega \hat {E}(\omega, \vec k) e^{i(\omega t-\vec k \vec r)}, \label{A.1}
\end{equation}
together with the usual relation between the frequency and the wave vector:
\begin{equation}
\omega = c ||\vec k ||.  \label{A.2}
\end{equation}
Using these notations, the time integrated field area is given by:
\begin{equation}
\int_{- \infty}^{+ \infty}dt E(t, \vec r)= \int_{- \infty}^{+ \infty}dt \int d\vec k \int d\omega \hat {E}(\omega, \vec k) e^{i(\omega t-\vec k \vec r)},  \label{A.3}
\end{equation}
which, upon the inversion of the integration order, leads to:
\begin{equation}
\int_{- \infty}^{+ \infty}dt E(t, \vec r)=  \int d\vec k \int d\omega \hat {E}(\omega, \vec k) e^{-i\vec k \vec r}\int_{- \infty}^{+ \infty}dt e^{i\omega t}.  \label{A.4}
\end{equation}
The last summation on the right-hand-side is nothing but the Dirac function:
\begin{equation}
\int_{- \infty}^{+ \infty}dt e^{i \omega t}=  2 \pi \delta (\omega ).  \label{A.5}
\end{equation}
Equation (\ref{A.4}) can then be simplified after integration over time and frequency as:
\begin{equation}
\int_{- \infty}^{+ \infty}dt E(t, \vec r)=  2 \pi \int d\vec k \hat {E}(0, \vec k) e^{-i\vec k \vec r}.  \label{A.6}
\end{equation}
It is clear from Eq.(\ref{A.2}) that a null frequency $\omega = 0$ has as a consequence a null momentum ($\vec k = 0$). We finally obtain
\begin{equation}
\int_{- \infty}^{+ \infty}dt E(t, \vec r)=  2 \pi \hat {E}(0, 0).  \label{A.7}
\end{equation}
The physical interpretation of Eq.~(\ref{A.7}) is as follows. The term $\hat{E}(0,0)$ actually represents a non-oscillating ($\omega=0$), non-propagating ($\vec k=0$) dc Stark field. Assuming a non-zero value for such a field requires, from the corresponding Maxwell equation, the necessary existence of a spatial electronic charge  distribution. In particular, finite charges placed at finite distances may create the dc Stark field in consideration. One may also invoke a charge distribution set at infinite distance, as limiting asymptotic conditions. But, of course, within this hypothesis, a non-zero Stark field would necessitate an overall distributed infinite electric charge. Ultimately, in systems where electric charges are not introduced on purpose, a dc Stark field cannot be created, and
\begin{equation}
\int_{- \infty}^{+ \infty}dt E(t, \vec r)=  2 \pi \hat {E}(0, 0) =0.  \label{A.8}
\end{equation}
\subsection{Designing control fields with zero-area constraint}
The goal of this section is to show how zero-area control fields can be designed. The three proposed methods are based on the optimization of the parameters of a functional form associated with the control problem. We consider a quantum system interacting with an electromagnetic field whose dynamics is described by the following time-dependent Schr\"odinger equation:
\begin{equation}\label{eqschrodinger}
i\frac{\partial}{\partial t}|\psi(t)\rangle =(H_0+E(t)H_1)|\psi(t)\rangle = H(t)|\psi(t)\rangle,
\end{equation}
where $H_0$ and $H_1$ are the field-free Hamiltonian and the interaction term, respectively, and $E(t)$ the control field. The units used throughout this paper are atomic units. Let $|\psi_0\rangle$ be the initial state and $t_f$ the total duration of the control. The goal of the control problem is to maximize the expectation value  $\langle \psi(t_f)|O|\psi(t_f)\rangle$  of a given observable $O$ at time $t=t_f$.

The first approach consists in introducing a closed-form expression for the control field depending on a finite number of parameters denoted $\alpha_i$:
\begin{equation}
E(t)=\mathcal{E}(\{\alpha_i\},t),
\end{equation}
with $\alpha_i\in \mathcal{A}$. The ensemble $\mathcal{A}$ is chosen such that the constraint on the time-integrated area, $A(t_f)=\int_0^{t_f}E(t)dt=0$, is satisfied. The optimal values of the  parameters $\alpha_i$ are determined in a second step by using gradient \cite{skinner} or global optimization procedures such as genetic algorithms \cite{dion,leibscher}. This approach, which can be very efficient in some cases, is however highly dependent on the parametrization used to describe the control field.

A more general method is based on an extension of the optimal and local optimization algorithms, which enforces the zero-area constraint through the introduction of a Lagrange multiplier. Such algorithms are proposed and investigated in Ref. \cite{JMO}. For the sake of completeness of the paper, we briefly outline below the principles of the different optimization procedures and we refer the interested reader to our preceding work for technical details on the numerical algorithms \cite{JMO}.

The optimal control problem is defined through a cost functional $J^{oc}$:
\begin{equation}\label{cost}
\mathcal{J}^{oc}=\langle \psi(t_f)|O|\psi(t_f)\rangle-\lambda \int_0^{t_f}(E(t)-E_{ref}(t))^2/S(t)dt-\mu [\int_0^{t_f}E(t)dt]^2.
\end{equation}
which allows us to maximize the expectation value $\langle \psi(t_f)|O|\psi(t_f)\rangle$ at time $t=t_f$ of a given observable $O$, while penalizing the total energy of the control field and enforcing the zero-area constraint. The novelty of the computational scheme resides in the introduction of a new Lagrangian multiplier $\mu$ to account for the zero-area constraint. In Eq. (\ref{cost}), the positive parameters $\mu$ and $\lambda$, expressed in a.u., weight the different parts of $\mathcal{J^{\rm{oc}}}$ with respect to the expectation value of $O$, and  penalize the area of the field ($\mu$- term) and its energy ($\lambda$- term). Larger is $\mu$ and closer is the time-area of the control field to zero. In Eq. (\ref{cost}), $E_{ref}$ is a reference pulse and $S(t)$ an envelope shape, which can be chosen as $S(t)=\sin^2(\pi t/t_f)$. As usual, note that the function $S(t)$ is introduced in the cost functional in order to ensure the smooth switch on and off of the field at the beginning and at end of the control.
Starting from this new cost functional, it is straightforward to derive a standard iterative algorithm based on Krotov (as used in this work) \cite{somloi,reich,maday,zhu,zhu1,densite} or gradient procedure \cite{grapeino}. The basic ingredient of the optimization procedure is the definition, at each step, of a new control field. At step $k$, we get:
\begin{equation}
E_{k+1}=E_k+\frac{S(t)\Im [\langle \chi_k |H_1|\psi_{k+1} \rangle ]}{2\lambda}-\frac{\mu}{\lambda}S(t)A_k,
\label{eq:newfield}
\end{equation}
where $|\psi_k(t)\rangle$ is the state of the system at the $k$th iteration, $|\chi_k (t)\rangle$ the adjoint state  obtained from backward propagation of the target $O|\psi(t_f) \rangle$ taken as an initial state for Eq.~(\ref{eqschrodinger}), $E_k$ and $E_{k+1}$ the control fields at steps $k$ and $k+1$, respectively, and $A_k$ the corresponding time-integrated area. Only the last term of the right-hand side of Eq. (\ref{eq:newfield}) is different from a standard procedure.

The local control theory (LCT) \cite{revuealessandro,engel,sugawara,wang} can also be extended along the same lines by considering the following Lyapunov function which accounts for the zero-area constraint:
\begin{equation}
\mathcal{J^{\rm{lc}}}(t)=\langle \psi(t)|O(t)|\psi(t)\rangle -\mu A(t)^2,
\label{eq:Jlc}
\end{equation}
where $O$ is any operator such that
\begin{equation}
i\frac{d}{d t}O(t)=[H_0,O(t)].
\label{eq:comutO}
\end{equation}
To ensure the monotonic increase of $\mathcal{J^{\rm{lc}}}$ at any time $t$, i.e. $\dot{\mathcal{J^{\rm{lc}}}}>0$, the control field is defined as follows:
\begin{equation}
E(t)=\eta \big(-i\langle\psi(t)|[O,H_1]|\psi(t)\rangle-2\mu A(t)\big),
\end{equation}
where $\eta$ is a positive parameter used to limit the amplitude of $E(t)$. We will also use in the following the parameter $\tilde{\mu}=\eta\mu$. Finally, note that the two proposed optimization procedures can only reduce the time-integrated area without completely removing it. A filtering process can further be  used to accurately design a zero-area field.
\section{Control of molecular orientation}\label{sec3}
This section is devoted to the control of the orientation dynamics of polar diatomic molecules \cite{friedrich,stapelfeldt,seideman,averbukh,salomon,lapertthz,tehini,pureorientation,mixedorientation,tehini2} with the constraint of zero-area fields. The two different approaches introduced in Sec. \ref{sec2} will be considered and discussed. We introduce a family of pulses characterized by a closed-loop expression depending on two parameters, which can be adjusted to enhance the degree of orientation. A second option is based on the optimization algorithms of Sec. \ref{sec2} which enforce the zero-area constraint through a Lagrange multiplier. No analytic expression of the optimal field can be derived in this case.
\subsection{Description of the model}
We consider a molecule described in a rigid-rotor approximation interacting with $E(t)$, a linearly polarized electromagnetic pulse along the $z$- axis of the laboratory frame. The electric field is assumed to be in the THz regime. The CO molecule in its ground vibronic state is taken as an illustrative example. At zero temperature, the dynamics is governed by the time-dependent Schr\"odinger equation (\ref{eqschrodinger}) where, in a linear approximation, $H_0=BJ^2$ and $H_1=-d\cos\theta$. The parameter $B$ is the rotational constant of the molecule, $J^2$ the angular momentum operator and $d$ the molecular permanent dipole moment. The spatial position of the diatomic molecule is given in the laboratory frame by its spherical coordinates $(\theta,\phi)$, $\theta$ being the angle between the molecular axis and the polarization vector, and $\phi$ the corresponding azimuthal angle. The Hilbert space associated with the dynamical system is spanned by the spherical harmonics $|j,m\rangle$, with $0\leq j$ and $-j\leq m \leq j$. We also recall that, due to the cylindrical symmetry of the problem, the projection $m$ of the total angular momentum on the field polarization axis is a good quantum number, so that $H(t)$ does not depend on the angle $\phi$. When the laser is switched on at $t=0$, the initial condition is given by $|\psi_0\rangle =|0,0\rangle$. The expectation value $\langle \cos\theta \rangle (t)=\langle \psi(t) |\cos\theta |\psi(t)\rangle $ is usually taken as a quantitative measure of orientation.

In the non-zero temperature case, the time evolution of the molecular system is described by the density operator $\rho(t)$ solution of the von Neumann equation:
\begin{equation}
i\frac{\partial \rho(t)}{\partial t}=[H(t),\rho(t)].
\end{equation}
The initial condition is a Boltzmann distribution at temperature $T$ and the degree of orientation is given by the expectation value expressed in terms of $\rho(t)$: $\langle \cos\theta \rangle (t)=\textrm{Tr} [\rho(t)\cos\theta]$. Numerical values of the molecular parameters are taken as $B=1.9312$ cm$^{-1}$ and $d=0.044$ a.u. The rotational period, $T_{per}$, is of the order of 8.64 ps or $3.57\times 10^5$ a.u. In the numerical computations, the maximum number of $j$- levels populated is of the order of 12.
\subsection{Control at zero temperature}
We first consider a class of pulses characterized by a closed-from expression in order to maximize the molecular orientation of the CO molecule at zero temperature. The electric field of this family can be written as follows \cite{lapertthz}:
\begin{eqnarray}\label{family}
& & E(t)=E_0\cos^2(\pi \frac{t}{\delta})\sin(2\pi f t),~t\in [-\delta/2,\delta/2]\\
& & E(t)=0,~t\notin [-\delta/2,\delta/2], \nonumber
\end{eqnarray}
with $E_0$ the pulse amplitude, $\delta$ its duration and $f$ its carrier wave frequency. It is straightforward to check that for any values of $\delta$ and $f$, $E(t)$ has actually a zero time-integrated area. The choice of this
form is motivated by the experimental feasibility of such pulses in the THz regime \cite{THZ1,THZ2,THZ3,THZ4,THZcontrol1,THZcontrol2}. We start our analysis by a general study of the degree of orientation that can be achieved within this class of fields. The field amplitude is fixed and corresponds to a peak intensity of 20 TW/cm$^2$, while the parameters $\delta$ and $f$ belong to the intervals $[0.12\times T_{per},0.25\times T_{per}]$ and $[0.5,3]$ THz, respectively. The maximum orientation achieved during the field-free evolution is indicated in Fig. 1 by a circle. We observe that the global degree of orientation is generally low, except for a zone of high orientation around $f=1$ THz and $\delta=0.15\times T_{per}$. The maximum of $|\langle \cos\theta\rangle |$ obtained is of the order of $0.88$ for $f=0.7$ THz and $\delta=0.14\times T_{per}$. The size of the high orientation region shows the robustness of the control pulse with respect to experimental imperfections while setting the parameters $\delta$ and $f$.
\begin{figure}[tb]
  \centering
  \includegraphics[width=0.8\linewidth]{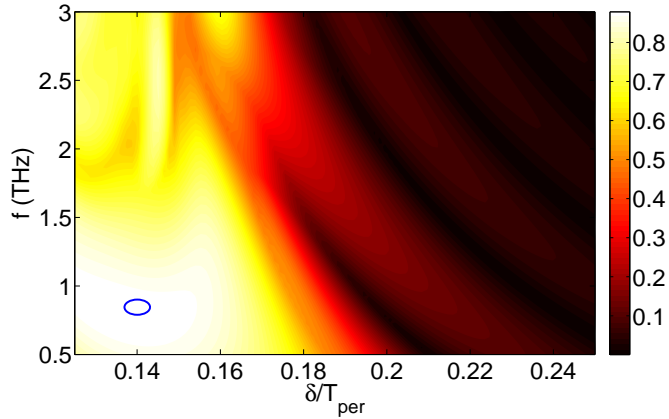}
  \caption{(Color online)
    Orientation efficiency $|\langle \cos\theta\rangle |$ (vertical color code) for the CO molecule as a function of the parameters $f$ and $\delta$ defined in Eq.(\ref{family}). The circle indicates the pulse for which the maximum degree of orientation is achieved.}
  \label{fig1}
\end{figure}

\begin{figure}[tb]
  \centering
  \includegraphics[width=0.75\linewidth]{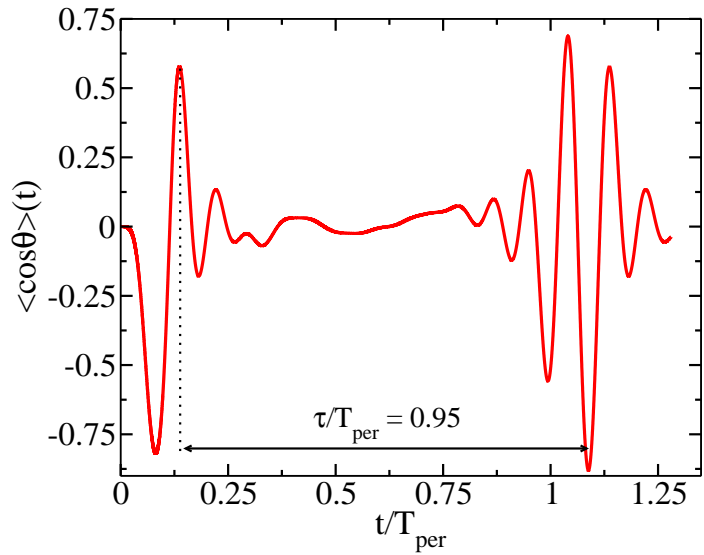}
  \caption{(Color online) Time evolution of $\langle\cos\theta\rangle$ induced by the guess field.
  The time $\tau$ is given by the relation $\tau \approx 0.95\times T_\mathrm{per}$.}
  \label{fig:evol_cos_i}
\end{figure}

\begin{figure}[tb]
  \centering
  \includegraphics[width=0.9\linewidth]{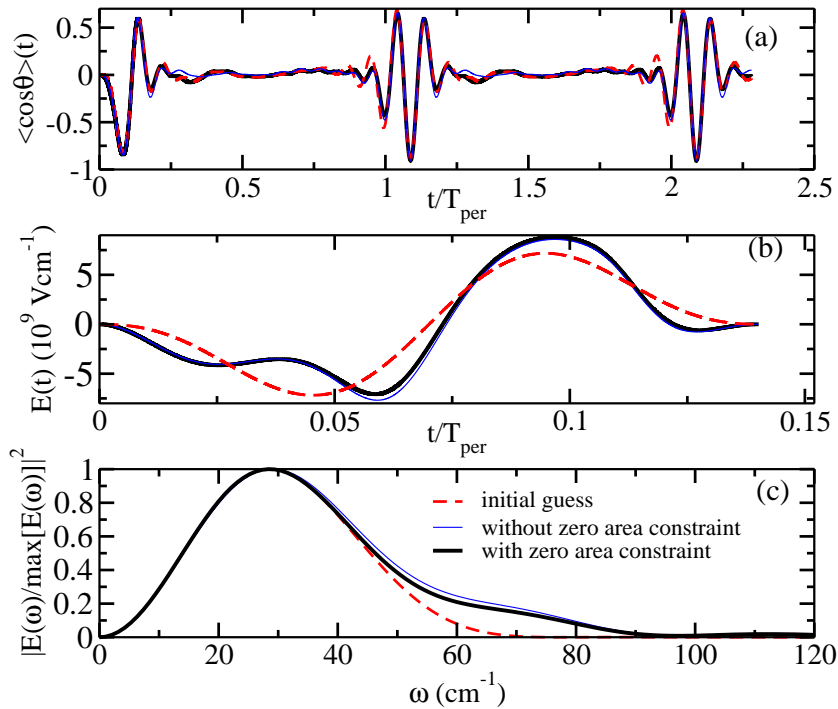}
  \caption{(Color online)
    (a) Time evolution of $\langle\cos\theta\rangle$ for $\mu=0$ (optimization without zero-area - solid thin blue line) and
  for $\mu \neq 0$ (solid thick black line - $\mu =  1.0\,10^{-4}$ a.u.).
The red dashed line represents the orientation dynamics induced by the initial guess field.
The panels (b) and (c) display the corresponding control fields and their Fourier transform.}
  \label{fig:JT_area}
\end{figure}

In a second step of the investigation, we use the best control field derived from the results of Fig. \ref{fig1} as a guess field for the optimal control algorithm. To be applied, this algorithm requires the definition of a target state. Here, we introduce the target operator $T$ defined as
\begin{equation}
  T = e^{-iH_0\tau}\cos\theta e^{iH_0\tau},
  \label{eq:tar_op}
\end{equation}
where $\tau$ is taken as the delay between the end of the guess pulse and the time where
$|\langle \cos\theta \rangle |(t)$ reaches its first maximum during the field-free evolution of the
system. Figure \ref{fig:evol_cos_i} displays the time evolution of $\langle \cos\theta \rangle (t)$
and gives the definition of the time $\tau$. This figure also illustrates how $\tau$ is chosen.
In numerical calculations, the intensity of the guess field
and the parameter $\lambda$  are fixed to  20 TW/cm$^{2}$ and 1, respectively, for both
optimizations, with and
without zero-area constraint. The delay $\tau$ in Eq.~(\ref{eq:tar_op}) is set to $0.95\times T_\mathrm{per}$.
The dynamics under the optimized fields is shown in
Fig.~\ref{fig:JT_area}(a), comparing the effect of standard and zero-area constraint algorithms.
Figure \ref{fig:JT_area}(b) shows the guess and the
 optimized fields with and without the zero-area constraint. Note that the global
shape of the two optimized fields is similar to the guess field, except for a small oscillatory behavior.
The optimized pulse without the zero-area constraint leads to higher orientation
($\max |\langle\cos\theta\rangle| \approx 0.906$)
than the pulse with the zero-area constraint ($\max |\langle\cos\theta\rangle| \approx 0.904$), but the time-integrated
area is divided by a factor 60 in the second case, going from -20.57 a.u. to -0.36 a.u. advocating for experimental feasibility. The very good orientation achieved
demonstrates the efficiency of the optimal control algorithms, even if the area of the field is no more strictly zero. As expected, we observe in Fig. \ref{fig:JT_area}(c) that the Fourier transform of the three fields
is equal or nearly equal to 0 at $\omega =0$. Since the optimized fields are very close to the guess
pulse, their  Fourier transforms show similar features. However,
the optimization slightly shifts the Fourier transform towards the high frequencies.
\subsection{Control at non zero temperature}
\begin{figure}[tb]
  \centering
  \includegraphics[width=0.9\linewidth]{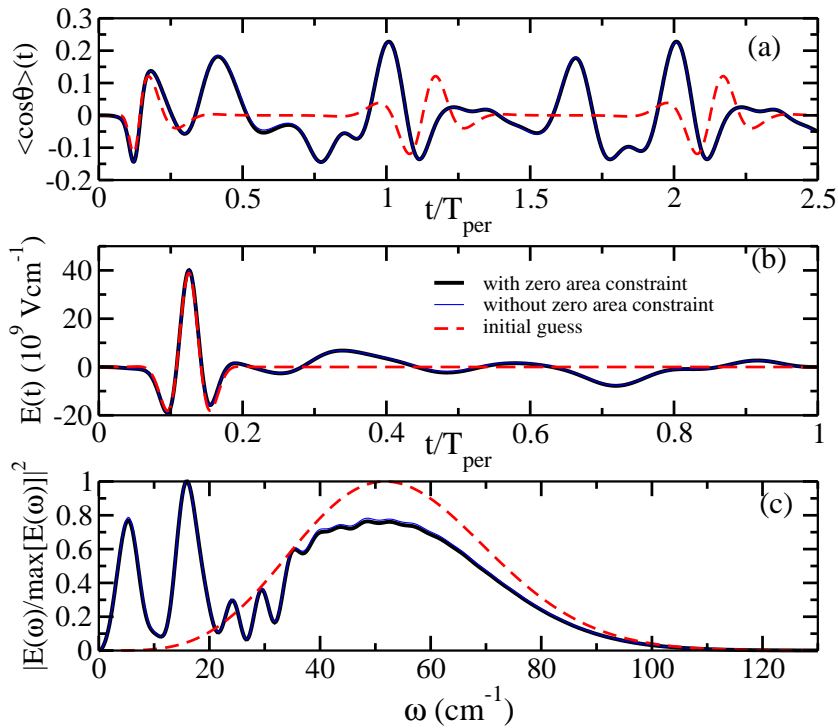}
  \caption{(Color online)
    Same as Fig. \ref{fig:JT_area} but for the non-zero temperature case.}
  \label{fig:JT_area_T30K}
\end{figure}

The efficiency  of the zero-area constraint algorithm is also checked for the CO molecule at non zero temperature.
This is a much more difficult task than controlling the orientation at zero temperature.
We discuss only the case of a long optimization time,
$t_f = T_\mathrm{per}$. We have considered as initial field a closed-form expression with zero area, which can be written as the sum of three Hermite polynomials (see Fig. \ref{fig:JT_area_T30K}(b) for a representation of this pulse). We have changed the maximum intensity from 20 TW/cm$^2$ to 2 TW/cm$^2$. The parameters $\lambda$ and  $\mu$ are  set to 20 a.u. and $1.8 \,10^{-4}$ a.u. respectively.
The temperature is fixed to 30 K. Figure \ref{fig:JT_area_T30K}(a) displays the time evolution of the expectation value of $\cos\theta$ induced by the guess and optimized fields. The  dynamics  under the optimized fields have similar features, but are very distinct from the one induced by the guess field.
Figure \ref{fig:JT_area_T30K}(b) compares the corresponding optimized fields together with the guess field.
As could be expected, the optimized fields show similar features.
In this example, the area of the optimized field with the zero-area constraint is two orders of magnitude
smaller than the one obtained without the zero-area constraint.
This is a remarkable result since the time-integrated area is largely reduced while preserving a satisfactory
orientation of the order of 0.2. The price to pay for increasing the final degree of orientation can be seen
in the Fourier transform of the optimized pulses, which have a much more complicated structure with an oscillatory
behavior at low frequency. These additional low frequencies found by the
  algorithm correspond to the  slow oscillations of the optimized fields which appear after
$t/T_{\mathrm{per}}>0.25$. By filtering out such oscillations, we have checked that this oscillatory behavior is essential to produce a high degree of orientation. Following Ref. \cite{henriksen}, we observe that the low frequency distribution of the optimized field coincides with the rotational resonance frequencies. This suggests an interpretation for the origin of the oscillatory behavior and a possible control mechanism based on the excitation of these different frequencies.
\section{Control of molecular photofragmentation}\label{sec4}
Another important application of local and optimal control strategies is illustrated on molecular photodissociation.
\subsection{Model system}
\begin{figure}[h]
 \centering
 \includegraphics[width=1\linewidth]{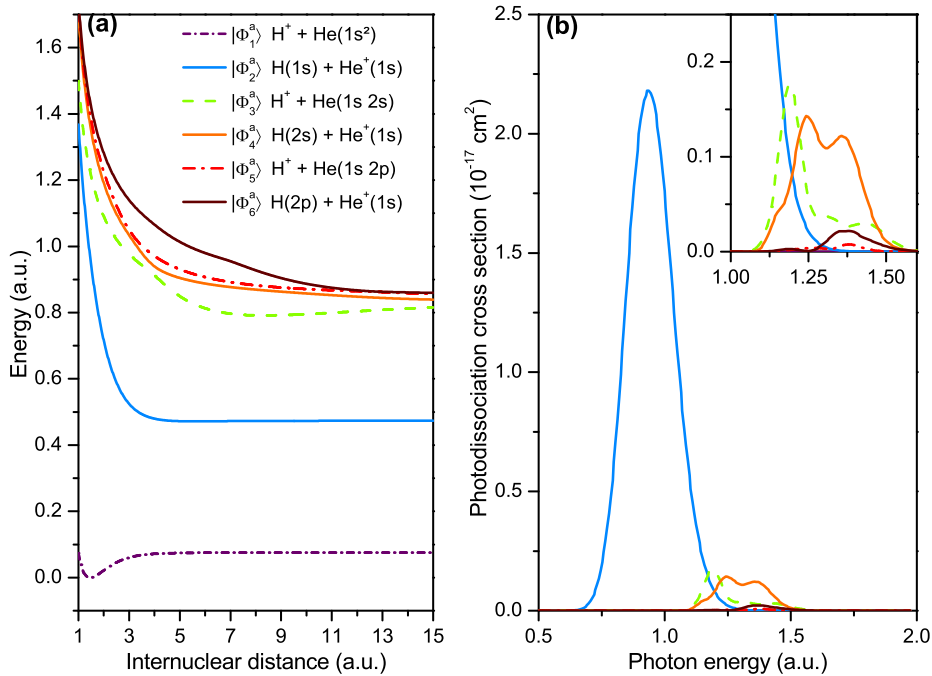}
 \caption{(Color online)
(a) Adiabatic potential energy curves of the $^1\Sigma$ states of HeH$^+$ leading to fragments in the $n=2$ shell. The target states are: H$^+$ + He(1s2s) (green dashed line), H$^+$ + He(1s2p) (red long dashes dots).
(b) Partial photodissociation cross sections together with an enlargement given as inset. The legend is the same as in panel (a).}
      \label{fig:HeH_pot}
\end{figure}

Due to the short duration of the pulses (as compared with the rotational period) a frozen rotation approximation is valid. In addition, the molecule is assumed to be pre-aligned along the $z$-direction of the laboratory frame. Therefore, the diatomic system is described by its reduced mass $m$ and the internuclear distance $R$. We aim at controlling the photodissociation of HeH$^+$ through the singlet $^1\Sigma$ excited states leading to He$^*$ fragment in the $n=2$ shell. We shall consider only parallel transitions among the singlet  $^1\Sigma$ states
induced by the $d_z$ dipole operator assuming the internuclear axis pointing along the $z$-direction. The adiabatic potential energy curves, the radial nonadiabatic couplings
$F_{ij}=\langle \Phi^a_i|\partial/\partial R|\Phi^a_j\rangle$ and the adiabatic transition elements $M^a_{ij}(R)$ of the $d_z$ dipole operator have been computed in Ref.~\cite{HeH}. Fig.~\ref{fig:HeH_pot} (a) displays the adiabatic potential energy curves. The partial photodissociation cross sections computed in Ref.~\cite{Sodoga} are displayed in Fig.~\ref{fig:HeH_pot} (b).

Dynamics is performed in the diabatic representation obtained from the adiabatic-to-diabatic transformation matrix $D$ which has been derived by integrating $\partial D/ \partial R +FD = 0$ from the
asymptotic region where both representations coincide. The total Hamiltonian introduced in Eq.~(\ref{eqschrodinger}) involves here
\begin{equation}
H_0 = \sum_{i,j=1}^N |\Phi^d_{i}\rangle  (T\delta_{ij}+V^d_{ij}(R)) \langle \Phi^d_{j} |
\label{eq:hamHeH+}
\end{equation}
and
\begin{equation}
H_1 = - \sum_{i,j=1}^N |\Phi^d_{i}\rangle M^d_{ij}(R) \langle \Phi^d_{j} |
\end{equation}
where $V^d_{ij}(R)$ are the potential matrix elements as obtained by diabatization of the adiabatic potential energies using the transformation matrix $D$. The parameter $N$ is the number of diabatic electronic states $| \Phi^d_{i} \rangle$ under consideration
and $T = - (1/2m )\partial^2 /\partial R^2  $.   $M^d_{ij}(R)$ are the
diabatized dipole transition matrix elements. The initial state ${|\psi_0 \rangle}$ is the lowest vibrational state of the ground electronic adiabatic state.

\subsection{Dissociation control}
\begin{figure}[h]
 \centering
 \includegraphics[width=1\linewidth]{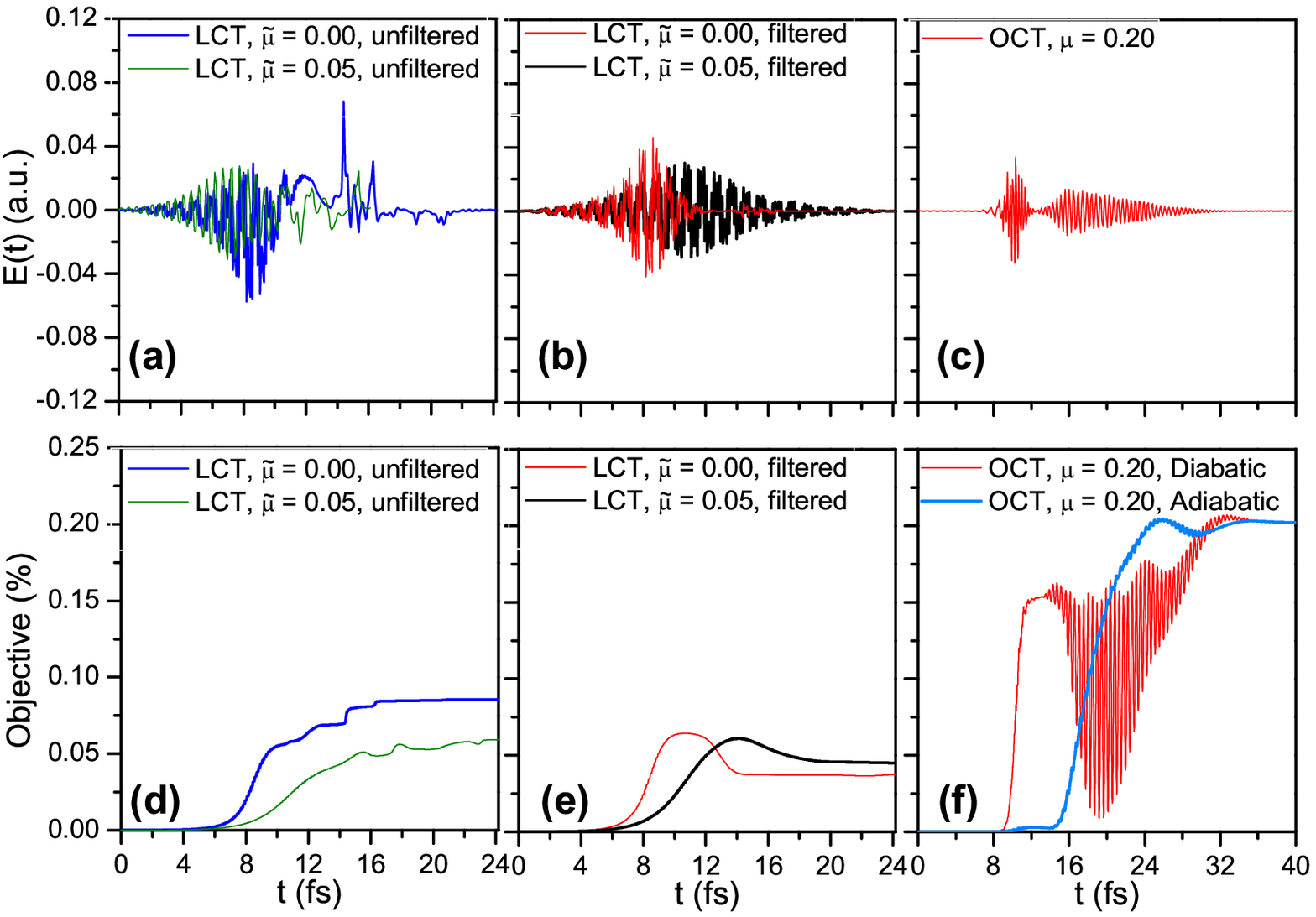}
 \caption{(Color online)
 Upper panels: Fields obtained by the local and optimal control of the photodissociation. (a) LCT without filtering of low frequencies, blue thick line: without constraint $\tilde{\mu} =0$, green thin line: with zero-area constraint $\tilde{\mu} =0.05$ a.u.,  (b) LCT with residual filtering, red thin curve: $\tilde{\mu} =0$ a.u., black thick curve: $\tilde{\mu} = 0.05$ a.u.. (c) OCT starting from the local field after filtration (red thin line of panel (b)),  red thin line: $\mu = 0.2$ a.u.
Lower panels: evolution of the objectives during the control for the different strategies. (d) The LCT objective is the population in the selected scattering states, blue thick line: without constraint $\tilde{\mu} =0$, green thin line: with zero-area constraint $\tilde{\mu} =0.05$ a.u., (e) Population in the adiabatic target states during the propagation with the filtered LCT pulses, red thin line: $\tilde{\mu} =0$ a.u., black thick line: $\tilde{\mu} = 0.05$ a.u. (f) Population in the target states during the propagation with the OCT pulse, red thin curve: diabatic representation, blue thick line: adiabatic representation. The asymptotic values give the He$^*(2s,2p)$ yields. }
      \label{fig:HeH_field}
\end{figure}

The goal is to enhance the yield in He$^*$ in the $n$= 2 shell through the dissociation channels leading  to He$^*(2s)$, or to He$^*(2p)$  \cite{LocalHeH}. These two target asymptotic states are denoted by $| \Phi^a_{3} \rangle$ (Fig.~\ref{fig:HeH_pot}, green dashed curve) and $| \Phi^a_{5} \rangle$ (Fig.~\ref{fig:HeH_pot}, red long dashes dots). In a first attempt we use a zero area Gaussian pulse with a carrier frequency chosen from the photodissociation cross section maximum.
We limit the total integrated intensity for further comparison with the OCT strategy. The yields remain very weak, of the order of 3 \% only, close to the value predicted by the fragmentation cross section. We then examine the efficiency of the zero-area constraint in both local and optimal control approaches. The local control can be considered as a very interesting first step before using OCT. In the presence of nonadiabatic interactions, the operator $O$ referred to in Eq.~(\ref{eq:Jlc}) requires to be chosen carefully since it has to commute with the field-free Hamiltonian $H_0$ (see Eq.~(\ref{eq:comutO})). The projectors on either adiabatic or diabatic states are thus not appropriate in LCT since they do not commute with this Hamiltonian due to kinetic or potential couplings, respectively \cite{LocalHeH}. This crucial problem can be overcome by using projectors on eigenstates of $H_0$, i.e. on scattering states correlating with the controlled exit channels. In this example, the  operator $O$ takes the form:
\begin{equation}
O  =  \sum_{p \in \mathcal{S}} \int dk |\xi_p^-(k)\rangle \langle \xi_p^-(k) |, \label{newproj}
\end{equation}
where $\mathcal{S}$ represents the two channels leading to the target He$^*(2s,2p)$ fragments, the objective being $\langle \psi(t)|O|\psi(t)\rangle$. The local control field now reads
\begin{equation}
E(t)  =   \eta \Im [( \sum_{p \in \mathcal{S}} \int dk \langle \psi(t) | \xi_p^-(k)\rangle \langle \xi_p^-(k) |d_z | \psi(t) \rangle )] -2 \tilde{\mu} A(t),
\end{equation}
involving two adjustable parameters $\eta$ and $\tilde{\mu}$. The ingoing scattering states $|\xi_p^-(k)\rangle$ are estimated using a time-dependent approach based on M{\o}ller-operators
\cite{tannorbook}
defined by
\begin{equation}
|\xi_p^-(k)\rangle  =  \lim_{t \rightarrow \infty} e^{i H_0 t} e^{-i H_f t} |p,k\rangle,
\end{equation}
where $|p,k \rangle$ is the outgoing plane wave in channel $p$ with energy $k^2/2 m$
and $H_f$ is the Hamiltonian operator of the fragments where all couplings have vanished asymptotically. This control strategy remains local in time but can preemptively account for nonadiabatic transitions that occur later in the dynamics. The photodissociation cross section \cite{Sodoga} shows that there is no spectral range where
the He$^*(2s,2p)$ dissociation channels dominate (see Fig.~\ref{fig:HeH_pot}(b)). The local control without any constraint ($\tilde{\mu} = 0$)  finds a very complicated electric field which
begins by a regular oscillatory pattern followed after 10 fs by a complex, positive real component shape whose area is obviously not zero (see blue thick curve in Fig.~\ref{fig:HeH_field}(a)). This erratic positive structure found in the LCT field can be interpreted as a Stark field. We therefore choose this example to check the efficiency of the zero-area constraint algorithm.

We first use the zero-area algorithm with $ \eta = 4.2$ a.u. and different values of $\tilde{\mu}$ (some examples are given in Ref.~\cite{JMO}). Figure \ref{fig:HeH_field}(a) (green thin curve) shows the pulse  for $\tilde{\mu}$ = 0.05 a.u. The algorithm efficiently reduces the Stark structure without completely removing it. The average objective (Eq.~(\ref{newproj})) for the two cases without (blue thick curve) and with the area constraint (green thin curve) are displayed in Fig. \ref{fig:HeH_field}(d). The objective is divided by about 2/3. As shown in Ref. \cite{JMO}, increasing $\tilde{\mu}$ to still reduce the Stark component decreases the objective so that a compromise has to be found. A complementary brute force strategy consists in removing the main part of the Stark component by filtration of near-zero frequencies. Starting from the initial LCT pulse, this already provides a large correction to the non vanishing area. The field after filtration of the low frequencies is shown in Fig. \ref{fig:HeH_field}(b) (red thin curve). To estimate the efficiency of the filtered pulse, we show the occupation of the two target adiabatic channels during the propagation in Fig. \ref{fig:HeH_field}(e) (red thin curve).  The final value of 3.75$\%$ is notably lower than the asymptotic value of the local objective 8.55 $\%$ (blue thick curve in Fig. \ref{fig:HeH_field}(d)). Figure \ref{fig:HeH_field}(b) (black thick curve) shows the pulse with $\tilde{\mu}$ = 0.05 a.u. after a subsequent filtering of the low-frequency components (compare with the green thin curve in Fig. \ref{fig:HeH_field}(a)). The  resulting regular profile confirms the efficiency of this mixed strategy. The population in the selected adiabatic channels with this filtered pulse is the black thick curve in Fig. \ref{fig:HeH_field}(e). The price to pay for reducing the pulse area is always a decrease of the target yield, but the best compromise is obtained when using both area constraint algorithm and residual filtering.
We first use the zero-area algorithm with $ \eta = 4.2$ a.u. and different values of $\tilde{\mu}$ (some examples are given in ref.\cite{JMO}). Figure \ref{fig:HeH_field}(a) (green curve) shows the pulse  for $\tilde{\mu}$ = 0.05 a.u. The algorithm efficiently reduces the Stark structure without completely removing it. The average objective (Eq.~(\ref{newproj})) for the two cases without (blue thick curve) and with area constraint (green thin curve) are drawn in Fig.~\ref{fig:HeH_field}(d)). The objective is divided by about 2/3. As shown in ref.\cite{JMO}), increasing $\tilde{\mu}$ to still reduce the Stark component decreases the objective so that a compromise has to be found. A complementary brute force strategy consists in removing the main part of the Stark component by filtration of near-zero frequencies. Starting from the initial LCT pulse, this already provides a large correction to the non vanishing area. The field after filtration of the low frequencies is shown in Fig.~\ref{fig:HeH_field}(b) (red thin curve). To estimate the efficiency of the filtered pulse, we show the occupation of the two target adiabatic channels during the propagation in Fig.~\ref{fig:HeH_field} (e) (red thin curve).  The final value of 3.75$\%$ is notably lower than the asymptotic value of the local objective 8.55 $\%$ (blue curve in Fig.~\ref{fig:HeH_field} (d)). Figure \ref{fig:HeH_field}(b) (black thick curve) shows the pulse with $\tilde{\mu}$ = 0.05 a.u. after a subsequent filtering of the low-frequency components (compare with the green curve in Figure \ref{fig:HeH_field}(a)). The  resulting regular profile confirms the efficiency of the mixed method. The population in the selected adiabatic channels with this filtered pulse is the thick black curve in Fig.~\ref{fig:HeH_field} (e). The price to pay for reducing the pulse area is always a decrease of the target yield, but the best compromise is obtained when using both area constraint algorithm and residual filtering.

\begin{figure}[h]
 \centering
 \includegraphics[width=1\linewidth]{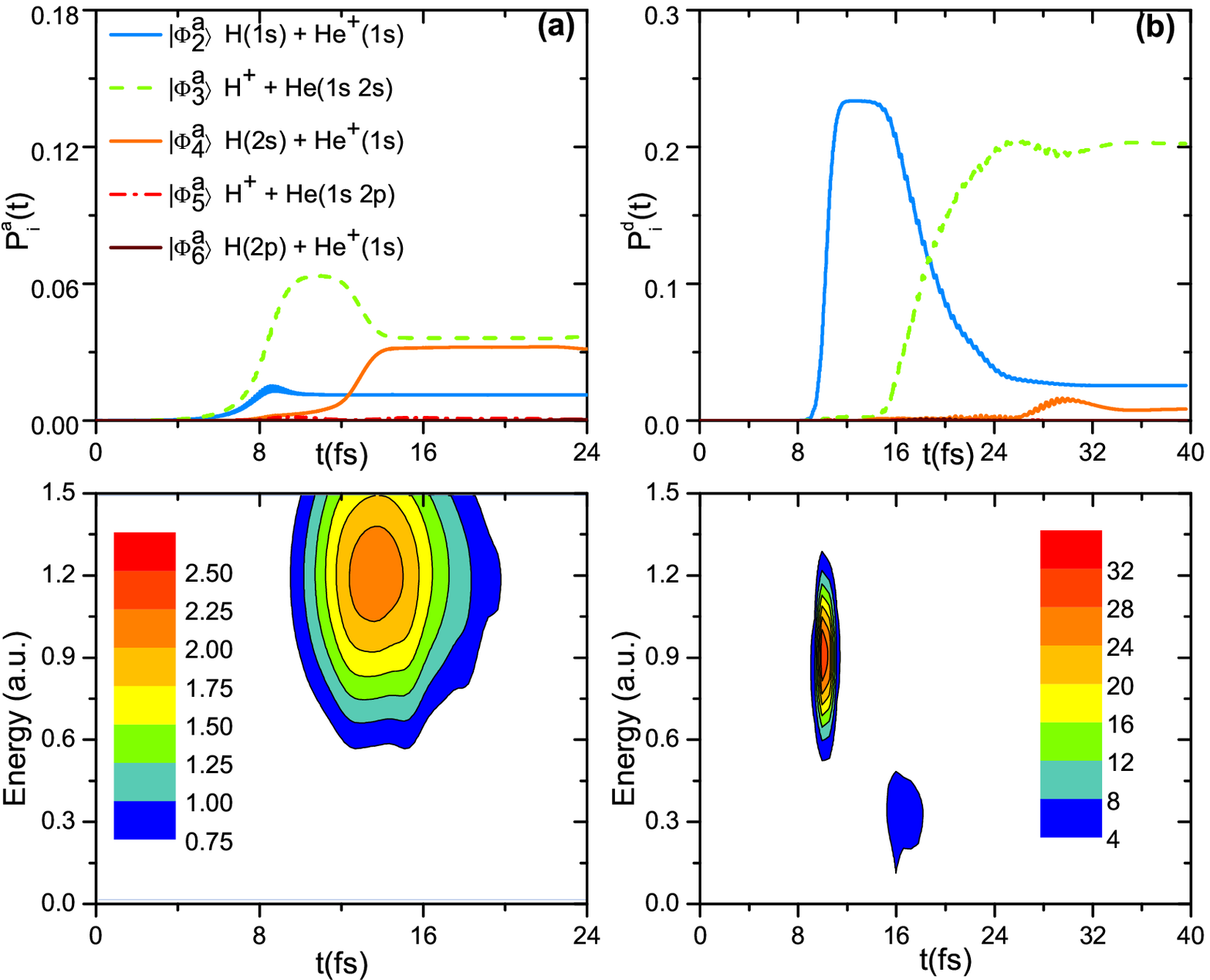}
 \caption{(Color online)
 Upper panels: Evolution of the adiabatic population in the different channels during the dissociation of HeH$^+$. The legend is the same for panels (a) and (b). The target states are: H$^+$ + He(1s2s) (green dahed line), H$^+$ + He(1s2p) (red dashes dots).
 Lower panels: Spectrograms of the fields. A color code with an arbitrary unit is given in each panel (c) or (d) to estimate the relative intensities.
 (a) Local control after filtration taken as guess field for OCT (red thin curve in Fig.~\ref{fig:HeH_field} (b))
 (b) Optimal control with $\mu = 0.2$ a.u. (red thin curve in Fig.~\ref{fig:HeH_field} (c))     }
      \label{fig:HeH_pop}
\end{figure}

In a second step, we explore the OCT strategy. Note that this procedure only refers to the zero-area algorithm without any subsequent filtration. The yield obtained with guess Gaussian fields increases only weakly while better results are obtained when the trial field is the LCT pulse. We choose the LCT pulse after filtration  (red thin curve in Fig.~\ref{fig:HeH_field}(b)) as guess field. Note that the OCT strategy only uses the zero-area algoritm without any subsequent filtration. The LCT field can be computed as long as the components in the excited states have some amplitude in the range covered by the initial ground vibrational state (roughly speaking, the Franck Condon region). This leads to a field vanishing after about 20 fs. In the global OCT strategy, the field is optimized on a time which may be longer. This opens the flexibility to exploit additional transitions towards the target states. We choose a final time $t_f = 40$ fs. The spatial grid is calibrated so that the target wave packet components do not reach the absorbing potential in the asymptotic region. The OCT objective is simply built here from the projector onto the diabatic states and the  operator $O$ takes the form:
\begin{equation}
O  =  \sum_{p\in \mathcal{S}} | \Phi^d_{p} \rangle \langle \Phi^d_{p} |. \label{newoctproj}
\end{equation}
As the objective is defined with the wave packet at the final time $t_f $, this corresponds to the required optimization of the decoupled scattering states. At each iteration, the final condition of the Lagrange multiplier is the asymptotic components in the target channels and no amplitude in all the other ones. The parameter $\lambda$ is chosen automatically by constraining the integrated intensity to 0.06 a.u. (see Ref.~\cite{gross}). The field corresponding to the best $\mu = 0.2$ a.u. is shown in Fig.~\ref{fig:HeH_field}(c) with a yield reaching $21.54\%$. As the simulation is performed in the diabatic basis set, the time evolution of the objective (see red thin curve in Fig.~\ref{fig:HeH_field}(f))  corresponds to the population in the sum of these two diabatic channels. The very strong oscillations reveal that the mechanism found by OCT in the last step is strongly non diabatic because the selected states are coupled with the other states during the process and decouple only at the end of the control. The mechanism is more simple in the adiabatic representation as can be seen by the evolution of the total population in the two adiabatic states correlated to the target fragments (blue thick curve in Fig.~\ref{fig:HeH_field}(f)). Fig.~\ref{fig:HeH_pop} compares the occupation of the adiabatic electronic states during the propagation with the guess field (panel (a)) and the best zero-area criterion optimal control pulse (panel (b)). The increase of the global target in OCT mainly comes from the enhancement of the $| \Phi^a_{3} \rangle$ (He*$2s$) (green dashed curve). OCT also reduces the unwanted transitions towards all other channels. The lower panels in Fig.~\ref{fig:HeH_pop} show the spectrograms of the filtered LCT and OCT fields. The main operating frequency of the local field corresponds to that predicted by the photodissociation cross section (at about 1.3 a.u.) for maximizing both channels. The inset in Fig.~\ref{fig:HeH_pot}(b) shows that this frequency corresponds to the maximum yield for the fragment $| \Phi^a_{3} \rangle$ (He*$2p$).  The additional mechanism due to a non optical Stark effect is thus suppressed by filtering very low frequencies. The OCT field first uses a low frequency component centered at about 0.8 a.u. This frequency favors of channel $| \Phi^a_{3} \rangle$ (He*$2s$) which explains the steep increase of that population and the vanishing influence of channel $| \Phi^a_{5} \rangle$ (He*$2p$). The $| \Phi^a_{2} \rangle$ channel is also more involved. After 20 fs, when the wave packet is out of the Franck Condon region, one observes a new mechanism proceeding via transitions  between the target and the $| \Phi^a_{4} \rangle$ and $| \Phi^a_{5} \rangle$  channels. These transitions require lower frequencies (about 0.3 a.u.) corresponding to the gap between the states (see fig.~\ref{fig:HeH_pot}(a)).

\section{Conclusion}\label{sec5}
After having discussed the physical origin of the time-integrated zero-area constraint on the laser control of molecular dynamics, we show that this fundamental requirement can be
included in the standard optimization computational schemes. A detailed description of the
dynamics achieved with such zero-area control fields is given and applied to two specific examples of molecular dynamics, namely the control of molecular orientation and that of molecular fragmentation.
Very encouraging results have been obtained even in the case of complicated quantum systems. In particular, we have derived for molecular orientation a closed-form expression of the control field depending only on two free parameters. The zero-area constraint is satisfied for any value of these parameters. At zero temperature, this approach reveals to be very efficient even when compared with the optimal solution. However, we have observed that the modified optimal control algorithm used in this work remains the best tool to handle more involved control problems, which cannot be solved by LCT or analytical fields with a sufficient efficiency.

This work and the possibility of including experimental constraints in optimal control algorithms
pave the way for future experimental implementations in quantum control. In other words, such results
help in bridging the gap between control theory and control experiments.\\ \\

\noindent\textbf{Acknowledgments}
S. V. acknowledges financial support from the Fonds de la Recherche Scientifique (FNRS) of Belgium. Financial
support from the Conseil R\'egional de Bourgogne and the
QUAINT coordination action (EC FET-Open) is gratefully
acknowledged by D. S. and M. N.. We thank the COST XLIC action.
O. A. acknowledges support from the European Union (Project No. ITN-2010-264951, CORINF).


\end{document}